\documentclass[twocolumn, prb, superscriptaddress, amsmath, amssymb,floatfix]{revtex4-2}
\usepackage{graphicx, nicefrac}
\usepackage[thinspace,thinqspace,amssymb]{SIunits}
\usepackage[version=4]{mhchem}
\usepackage{chemformula}
\usepackage{color,hyperref}
\usepackage{float}
\usepackage{ulem}
	\definecolor{LinkColor}{rgb}{0.45,0,0}
	\definecolor{UrlColor}{rgb}{0,0,0.45}
	\definecolor{CiteColor}{rgb}{0,0.45,0}
	\definecolor{olivegreen}{RGB}{0, 50.2, 0}
    \definecolor{mark}{RGB}{205,222,255}
	\hypersetup{%
	linkcolor=LinkColor,%
	filecolor=LinkColor,%
	menucolor=LinkColor,%
	citecolor=CiteColor,%
	urlcolor=UrlColor,%
	colorlinks=true%
	}

\renewcommand{\emph}{\textit}

\newcommand{\Mag}{\textbf{M}}
\newcommand{\XMDP}{\mbox{HAX-MDP}}
\newcommand{\HAXPES}{\mbox{HAXPES}}
\newcommand{\PETRA}{\mbox{PETRA~III}}
\newcommand{\TC}{$T_C$}
\newcommand{\TSM}{$T_{s\text{\textbf{M}}}$}

\newcommand{\STO}{SrTiO$_\text{3}$}

\begin{document}
 \setchemformula{kroeger-vink}

\title{2D synthetic ferrimagnets by magnetic proximity coupling}

	\author{Paul Rosenberger}
	\affiliation{Fachbereich Physik, Universit\"at Konstanz, 78457 Konstanz, Germany}
	\affiliation{Fakult\"at Physik, Technische Universit\"at Dortmund, 44221 Dortmund, Germany}
	
	\author{Moumita Kundu}
	\affiliation{Fachbereich Physik, Universit\"at Konstanz, 78457 Konstanz, Germany}

	\author{Andrei Gloskovskii}
	\affiliation{Deutsches Elektronen-Synchrotron DESY, Notkestrasse 85, 22607 Hamburg, Germany}
	
	\author{Christoph Schlueter}
	\affiliation{Deutsches Elektronen-Synchrotron DESY, Notkestrasse 85, 22607 Hamburg, Germany}
	
	\author{Ulrich Nowak}
	\affiliation{Fachbereich Physik, Universit\"at Konstanz, 78457 Konstanz, Germany}

	\author{Martina M\"uller}
	\email{martina.mueller@uni-konstanz.de}
	\affiliation{Fachbereich Physik, Universit\"at Konstanz, 78457 Konstanz, Germany}

	\date{\today{}}

\begin{abstract}
Proximity effects allow for the adjustment of magnetic properties in a physically elegant way. If two thin ferromagnetic (FM) films are brought into contact, electronic coupling alters their magnetic exchange interaction at their interface. For a low-\TC~rare-earth FM coupled to a 3d~transition metal FM, even room temperature magnetism is within reach. In addition, magnetic proximity coupling is particularly promising for increasing the magnetic order of metastable materials such as europium monoxide (EuO) beyond their bulk \TC, since neither the stoichiometry nor the insulating properties are modified.

We investigate the magnetic proximity effect at Fe/EuO and Co/EuO interfaces using hard X-ray photoelectron spectroscopy. By exciting the FM layers with circularly polarized light, magnetic dichroism is observed in angular dependence on the photoemission geometry. In this way, the depth-dependence of the  magnetic signal is determined element-specifically for the EuO and 3d~FM parts of the bilayers. In connection with atomistic spin dynamics simulations, the thickness of EuO layer is found to be crucial, indicating that the observed antiferromagnetic proximity coupling is a short-ranged and genuine interface phenomenon. This fact turns the bilayer into a strong synthetic ferrimagnet. The increase in magnetic order in EuO occurs in a finite spatial range and is therefore particularly strong in the 2D limit---a counterintuitive but very useful phenomenon for spin-based device applications.

\end{abstract}

\maketitle
	

	\section{Introduction}
	\label{sec:R3:introduction}
	Controlling the spin degree of freedom is a key aspect for emerging quantum information technologies~\cite{kim_2022}. Taking advantage of quantum phenomena, such as spin-dependent tunneling~\cite{mueller_2009,caspers_chemical_2011}, spin-Hall magneto-resistance~\cite{rosenberger_quantifying_2021}, confined quantum wells~\cite{prinz_quantum_2016} or two-dimensional electron systems~\cite{lomker_two-dimensional_2017}, ferromagnetic semiconductors are ideally suited templates for controlling and creating spin-polarized states. Unfortunately, these materials, such as the Europium monochalcogenides, suffer from notoriously low Curie temperatures of \TC $=69\,$K for EuO or even lower for EuS and EuSe~\cite{mcguire_ferromagnetic_1964, mueller_thickness_2009}. Therefore, the search for ways to adjust their magnetic ordering temperature attracted attention already decades ago~\cite{mcwhan_magnetic_1966,abd-elmeguid_onset_1990,fumagalli_exchange-induced_1998,altendorf_spectroscopic_2012}. While chemical doping obviously has a detrimental effect on the desired semiconductivity, proximity effects at the interface between two ferromagnets with higher and lower values of \TC)  prove to be a promising approach for enhancing magnetic order in the low-\TC~material~\cite{fumagalli_exchange-induced_1998,lewitz_proximity_2012,pappas_direct_2013,poulopoulos_induced_2014,goschew_verification_2016,verenas-paper}.\\	
	Interfaces of 3d~ferromagnets with EuO and its parent compound EuS have been investigated in several previous studies: In 1969, Ahn and Almasi reported already an antiferromagnetic~(AFM) type coupling between Fe and EuO ~\cite{ahn_coupling_1969}, and recently AFM coupling was demonstrated in the transient magnetization in Co/EuO bilayers~\cite{moenki_EuO_2023}. Later, several studies addressed the enhancement of magnetic order of EuS when it is interfaced with Fe, Ni or Co~\cite{fumagalli_exchange-induced_1998,lewitz_proximity_2012,pappas_direct_2013,poulopoulos_induced_2014,goschew_verification_2016,jensen_enhanced_1999}. While AFM coupling was reported in all cases, the enhancement of magnetic order was shown to be the more effective, the thinner the EuS is, suggesting that the enhancement of magnetic order is a localized effect. Compared to other, not even related material systems, a depth dependence, i. e. limited range, of the magnetic proximity effect was hypothesised to be the cause for these observations~\cite{carrico_phase_1992}.  To date, however, neither the microscopic details of the proximity effect itself nor the thickness dependence of the enhancement of magnetic order are known. For a better optimization of proximity-enhanced magnetic order, the investigation of the depth dependence of the magnetic ordering within these bilayers is therefore of crucial importance.
	In the present work, we study the depth-dependent magnetic moment of 3d~FM/EuO bilayers using an approach based on hard X-ray photoelectron spectroscopy~(HAXPES)~\cite{mueller_hard_2022}. In the following, we will refer to it as hard X-ray magnetic depth profiling (\XMDP). \XMDP~unites three features of photoelectron spectroscopy, i)~element selectivity, ii)~sensitivity to magnetism due to magnetic circular dichroism in photoemission (XMCD-PE), and iii)~bulk sensitivity in case hard X-rays are used, i. e. HAXPES. It is therefore an ideal tool for studying magnetic order \emph{at} and \emph{near} buried interfaces.
	
	Applying \XMDP~to 3d~FM/EuO bilayers reveals insights into the nature of the magnetic proximity effect at the interface. Furthermore, we use  atomistic spin dynamics simulations to study the depth-dependence of the proximity effect from a theoretical perspective. Comparing measurements with simulations we can show, that the strength of this proximity-induced interface magnetization depends not only on temperature but also on the thickness of the EuO layer. We can demonstrate, that, in the very thin limit of quasi-2D EuO layers, the interface magnetic order of EuO will survive for temperatures up to room temperature. Furthermore, the antiferromagnetic interface coupling in connection with the proximity effect converts our bilayers into a synthetic ferrimagnets with a compensation temperature which is, surprisingly, above the Curie temperature of pure bulk EuO.  
\section{Methods}

\subsection{Sample preparation}
\label{sec:R3:preparation}
We prepared Fe/EuO and Co/EuO heterostructures as well as EuO reference samples on TiO$_2$ terminated \STO:Nb substrates (Crystec GmbH) using molecular beam epitaxy (MBE). The base pressure of the oxide MBE system (at University of Konstanz) was \mbox{p$_\text{MBE}\,\leq\,2\,\times\,10^{-10}\,$mbar}. Prior to EuO growth, the substrates were annealed for 2h at $T_\text{S}\,=\,600^\circ$C in an oxygen atmosphere. For EuO synthesis, we made use of the redox growth process as described elsewhere~\cite{lomker_redox-controlled_2019, rosenberger_europium_2022}. Here, the sample temperature was $T_\text{S}\,=\,500^\circ$C. Eu metal was evaporated from a Knudsen cell with a rate of $r_\text{Eu}\,=\,0{.}13\,$\AA/s which was measured using a quartz crystal micro balance. The desired EuO thicknesses were $3\,$nm and $11\,$nm.\\
The EuO stoichiometry was subsequently confirmed by \emph{in situ} X-ray photoelectron spectroscopy (XPS).\\
Afterwards, we deposited the Fe (Co) overlayers of a thickness of $4\,$nm using e-beam evaporation at $T_S\,=\,$RT with rates of $r_\text{Fe,Co}\,\approx\,0{.}08\,$\AA/s.\\
Samples were then stored and transferred to the \HAXPES~spectroscopy instrument of beamline P22 at \PETRA~(DESY, Hamburg) inside an ultra-high vacuum suitcase at a pressure of $p_\text{SC}\,\leq\,3\,\times\,10^{-11}\,$mbar. At P22, the suitcase was attached to the end station's load lock, enabling a full \emph{in vacuo} sample handling from preparation to measurement.

\subsection{Setup}
\label{sec:R3:SetupMethod}
\begin{figure}[tbh]
        \centering
    	\hspace*{-1mm}
        \includegraphics[width=0.7\columnwidth,clip]{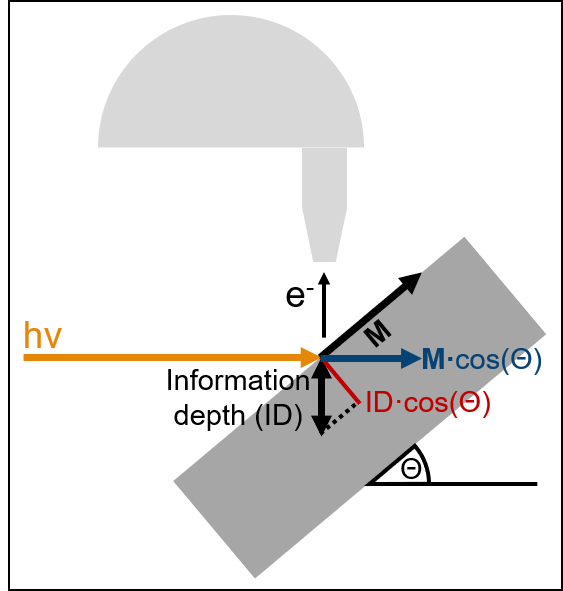}\quad
    	\caption{Sketch of the experimental setup for magnetic depth profiling by HAX-MDP. Photoelectrons are detected under an angle of $90^\circ$ relative to the photon beam incidence. The in-plane magnetized sample is tilted by a polar angle $\Theta$, yielding the depth dependence of the XMCD in photoemission and the scaling of its sensitivity on \Mag.}
        \label{fig:SketchSetup}
    \end{figure}

    \HAXPES~measurements were performed using the endstation of beamline P22 at \PETRA~(DESY, Hamburg)~\cite{P22_HAXPES_2019}. The setup provides a $90^\circ$ angle between photon beam direction and the SPECS Phoibos 225 hemispherical electron analyzer, see Fig.~\ref{fig:SketchSetup}. The photon energy was set to $6\,$keV, while the most strongly bound electrons under investigation (Eu~3d core level) have a binding energy $E_\text{B}\,\leq\,1{.}2\,$keV. Therefore, a significant depth sensitivity is guaranteed due to the effective attenuation length (EAL) of $\,\approx\,6{.}1\,$nm in EuO and $4{.}9\,$nm ($5{.}2\,$nm) in the Co (Fe) overlayer~\cite{jablonski_effective_2015,SESSA_2014}. In order to obtain a magnetic circular dichroism, the light helicity was switched between $\sigma^+$ and $\sigma^-$ using a phase retarder.

\subsection{Hard X-ray magnetic depth profiling}
\label{sec:R3:HAX-MDP}
        We investigated the element specific magnetism of the EuO film and the 3d~FM overlayer by means of X-ray magnetic circular dichroism (XMCD) in photoemission (PE), i.e. \mbox{XMCD-PE}. The general mechanism has been described in literature and has been validated for Eu(O)~\cite{starke_magnetic_2000,caspers_magnetic_2013}.
        The \mbox{XMCD-PE} is calculated as follows:\\ 
        First, a constant background was subtracted from the spectra recorded for the two light helicities $\sigma^+$ and $\sigma^-$, respectively. Next, both spectra were normalized to their respective integrals. The \mbox{XMCD-PE} was then calculated using
      \begin{align}
            \label{Eq_XMCD-PE}
            \text{XMCD-PE}(E)= 100\% \times \frac{I^+(E)-I^-(E)}{\text{max}[I^+(E)+I^-(E)]} \, .
        \end{align}
        In order to reduce noise effects leading to an overestimated \mbox{XMCD-PE}, a three point adjacent average smoothing was applied. The given absolute values of the (smoothed) \mbox{XMCD-PE} were finally obtained by taking the difference between its maximum and minimum, i.e. twice the amplitude.\\
        We measured the \mbox{XMCD-PE} signal under various polar (emission) angles $\Theta$ as sketched in Fig.~\ref{fig:SketchSetup}. That way, the \Mag-component parallel to the incident photon beam direction is probed. Hence, for a purely in-plane magnetized sample, the obtained \mbox{XMCD-PE} scales with $\cos{(\Theta)}$, while an out-of-plane component of \Mag~results in a contribution scaling with $\sin{(\Theta)}$. 
        Considering the energy dependent attenuation length of photoelectrons inside a solid, a $\cos{(\Theta)}$-scaling also applies to the effective information depth. This enables depth profiling of the element specific magnetic moment. Here, we apply \XMDP~to gain insights into the magnetic proximity effect at buried 3d~FM/EuO interfaces.
        
	\begin{figure}[tbh]
		\centering
		\hspace*{-1mm}
		\includegraphics[width=0.9\columnwidth,clip]{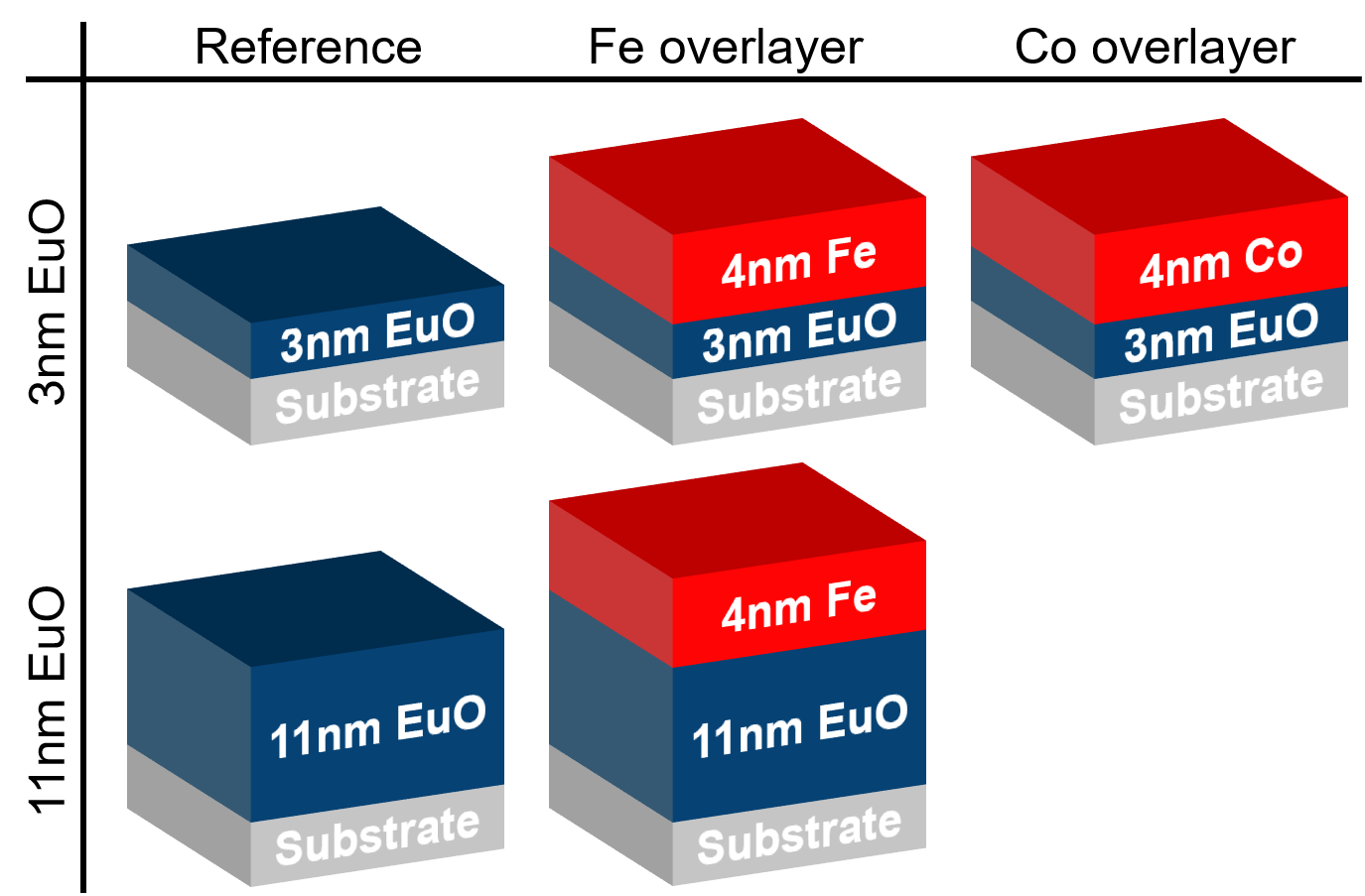}\quad
		\caption{Overview of the different samples used in this study. For both EuO film thicknesses, $3\,$nm and $11\,$nm, there were reference samples without metal overlayer and samples with a $4\,$nm Fe capping. Co as capping material was studied on thin EuO only.}
		\label{fig:SamplesOverview}
	\end{figure}	    
\subsection{Theoretical modeling}
For a deeper understanding of the  mechanisms which are responsible for the proximity effect as seen in the experiments we employ atomistic spin dynamics simulations. In this approach, the total atomic magnetic moment at each lattice site $i$ is $\mathbf{\mu}_{s,i} = \mu_{s,i}\mathbf{S}_i$, where $\mathbf{S}_i$ is a unit vector defining the direction of the magnetic moment. The whole bilayer is described via a  Hamiltonian of Heisenberg type,
\begin{align}
\label{1}
\mathcal{H}_{j} &= - \frac{1}{2}\sum_{j,k(NN)=1}^N \left[\mathbf{S}^T_j \mathcal{J}_{j,k} \mathbf{S}_k  \right] \notag \\    
&- \sum_{j=1}^N d_{j,4}	\left[ S_{j,x}^4 + S_{j,y}^4 + S_{j,z}^4 \right] - \sum_j \mu_{s,j} \mathbf{S}_j \cdot \mathbf{B_{ext}},
\end{align}
where, $\mathcal{J}_{j,k}$ is the exchange tensor which includes both, the exchange interaction (restricted to nearest neighbors (NN) where, $J_{EuO}=1.8797$ meV, $J_{Fe}=43.7457$ meV and $J_{Co}=33.336$ meV) and the shape anisotropy. The latter is modeled as an effective two-ion anisotropy with the  $z$ axis as hard axis, favoring the $x$-$y$-plane as easy plane.
$d_{j,4}$ is the cubic anisotropy which, along with the shape anisotropy, defines the energetically favored ground state of the spins. $\mathbf{B_{ext}}$ is the external magnetic field which is applied to align the magnetization along $x$-direction before increasing the temperature, similar to the experiments. 

EuO grows in a rock-salt structure but since we model just the magnetic ions, we focus solely on the $Eu^{+2}$ ions which are arranged on an FCC lattice, with a resulting cubic anisotropy having a value of $-9.3*10^{-3}$ meV \cite{EuO_anisotropy}. The cubic anisotropy being negative prefers the body diagonal direction, and competes with the shape anisotropy which prefers in-plane alignment. The shape anisotropy for these films is comparable due to the thin-film structure and the very high magnetic moment of EuO of $\approx 6.99\mu_B$ per atom. Over all, this results in a preferential easy-plane alignment of the spins. The Fe and Co layers are modelled in an BCC and FCC structure, respectively. In the experiments, however, the transition metals that are deposited on top of EuO are amorphous and the cubic anisotropy in those materials is practically zero. Their alignment is defined by the material with stronger magnetic moment, i.e., EuO or by the initially applied strong magnetic field that aligns the spins towards or opposite to the direction of the magnetic field. 

To get the time-evolution of spins we use the stochastic Landau-Lifshitz-Gilbert equation \cite{Gilbert_1955_ALF} as the equation of motion,
\begin{align}
\label{2}
\frac{\partial \mathbf{S_j}}{\partial t} &= - \frac{\gamma}{\mu_s (1+\alpha^2) }[\mathbf{S_{j}} \times \mathbf{H_{j}}+\alpha 	\mathbf{S_{j}} \times (\mathbf{S_{j}} \times \mathbf{H_{j}})],
\end{align}
where, $\gamma > 0$ is the gyromagnetic ratio, and $\alpha$ the Gilbert damping constant~\cite{Gilbert_1955_ALF}. The first term on the right describes the precession of the spins around the effective field, $\mathbf{H_j}$ and the second term describes how the spins relax towards the direction of the effective field. $\mathbf{H_j}$ can be expressed as
\begin{equation}\label{3}
	\mathbf{H_j} = -\frac{\partial \mathbf{\mathcal{H}_j}}{\partial \mathbf{S}_j} + \mathbf{\xi_j},
\end{equation}
where $\mathbf{\xi_j}$ is a Gaussian white noise with
\begin{equation}\label{4}
\langle {\xi_j}^\beta (t) {\xi_k}^\zeta(t') \rangle = \frac{2{\mu_j^s} \alpha k_{B}T}{\gamma_j} \delta_{jk} \delta_{\beta\zeta}\delta(t-t').          
\end{equation}
This noise describes thermal fluctuations  of the spin system due to the coupling to a heat bath, which is at temperature $k_B T$. The noise is uncorrelated in space and time and $\beta$ and $\zeta \in \left\lbrace x, y, z\right\rbrace $.
These differential equations are then solved using the stochastic Heun's method using a code developed in C++ and CUDA which allows us to run the simulations in graphical processing units for faster computation. The simulated system size is $128\times128\times16$ unit cells. This then scales with the number of atoms per unit cell depending on the lattice structure of the samples leading to about $10^6$ spins. The simulations are performed using the HPC cluster SCCKN of the University of Konstanz.
\section{RESULTS}
	    \label{sec:R3:ResultXMDPEUO}
		\begin{figure*}[ht!]
		\centering
		\hspace*{-1mm}
		\includegraphics[width=2.00\columnwidth,clip]{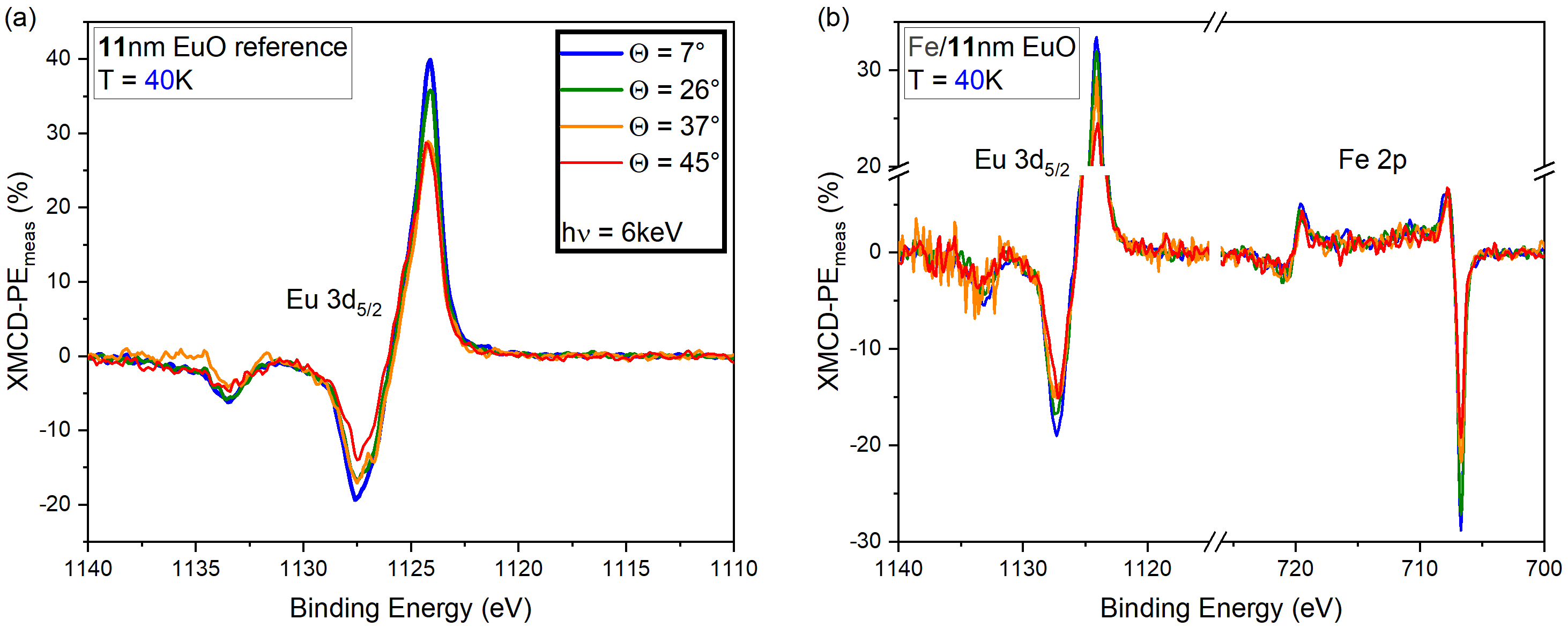}
		\caption{Strong XMCD-PE$_{\text{meas}}$ obtained under different polar angles $\Theta$ at $T_{\text{low}}$ from (a) a bare $11\,$nm EuO film and (b) an $11\,$nm EuO film with a $4\,$nm~Fe overlayer. As expected for the in-plane magnetized sample, for both Eu and Fe, the XMCD-PE$_{\text{meas}}$ decreases with increasing $\Theta$.} 
		\label{fig:Fe11nmEuO_Tlow_XMCDPE_allAngles}
	\end{figure*}
	Based on the \mbox{XMCD-PE} of the 3d$_\text{5/2}$ core level of Eu and 2p core level of the 3d~FMs, we present \XMDP~data of four samples: (i) an $11\,$nm thick bare EuO film, (ii) an $11\,$nm thick EuO film with a $4\,$nm Fe overlayer, (iii) a $3\,$nm thin EuO film with $4\,$nm Fe overlayer and (iv) a $3\,$nm thin EuO film with a $4\,$nm Co overlayer. For comparison, also the \mbox{XMCD-PE} of a bare $3\,$nm EuO film was recorded in (close to) normal emission~(NE) geometry. Sketches of the samples are shown in Fig.~\ref{fig:SamplesOverview}.\\
	Data were recorded at the lowest experimentally achievable temperature, \mbox{$T_{\text{low}}\,\approx\,40\,$K}, i.e. well below the bulk Curie temperature of EuO ($T_{\text{C}}\,=\,69\,$K) and at $T_{\text{high}}\,=\,80\,\text{K}\,>\, T_{C}$. After cooling down a sample mounted to the cryostat of the sample manipulator and prior to the measurements, the sample was magnetized in-plane by approaching a permanent magnet. The magnet was subsequently removed and measurements took place with the sample in the remanent state. Measurements were carried out under four polar angles, \mbox{$\Theta\,=\,7^\circ$}, $26^\circ$, $37^\circ$ and $45^\circ$.

\subsection{XMCD-PE in normal photoemission geometry}
\begin{figure*}[tbh]
		\centering
		\hspace*{-1mm}
		\includegraphics[width=2.00\columnwidth,clip]{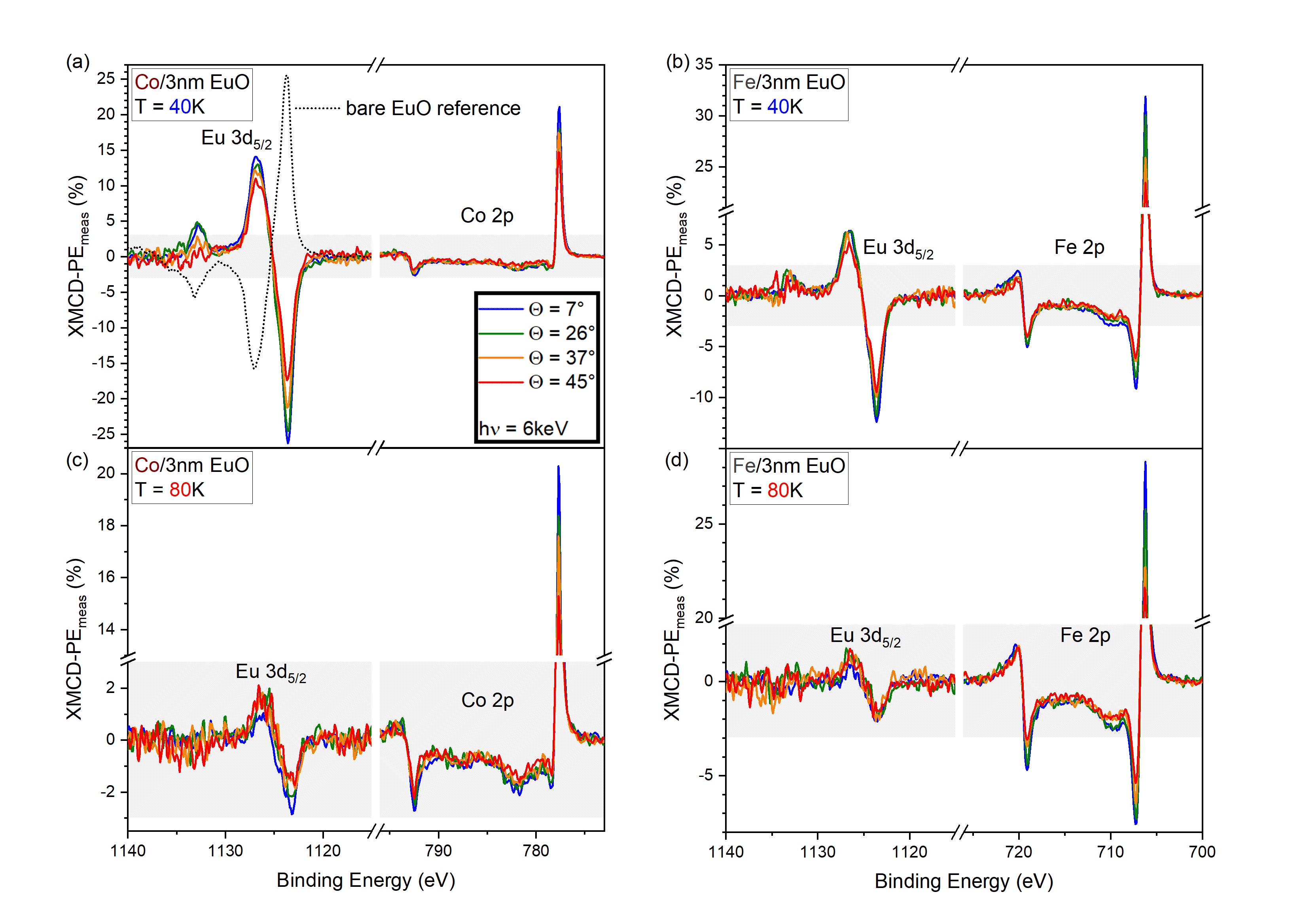}
		\caption{Angle dependent XMCD-PE$_{\text{meas}}$ of the Eu~3d$_{5/2}$ and Co (Fe) 2p core levels obtained from $3\,$nm thin EuO films capped with $4\,$nm of Co, (a), (c), and Fe, (b), (d) at $T_{\text{low}}$, (a), (b), and $T_{\text{high}}$, (c), (d). The dashed line in (a) is the reference signal of a bare $3\,$nm thin film of EuO in NE ($\Theta\,=\,3^\circ$). Note that for the sake of visibility the scaling of the y-axes varies. The grayish area serves as a guide to the eye, indicating the \mbox{XMCD-PE$_{\text{meas}}$} range from $-3$ to $+3$. For all samples, angles and temperature, a non-zero XMCD-PE$_{\text{meas}}$ of Eu is observed. Obviously, interfacing a thin EuO film with Co or Fe switches the magnetization direction with respect to the bare EuO film. At $T_{\text{low}}$ the XMCD-PE$_{\text{meas}}$ of Eu is larger for the film interfaced with Co than with Fe. The amplitudes of all XMCD-PE$_{\text{meas}}$ decrease with increasing $\Theta$.} 
		\label{fig:all3nmEuO_allT_XMCDPE_allAngles}
	\end{figure*}

	We first discuss the \mbox{XMCD-PE} of the samples recorded in NE and at $T_{\text{low}}$, where we initially assume the magnetization to be in-plane and along the axis of the magnetic field applied to magnetize the samples. Note that measurements were performed slightly off NE. Therefore, the \mbox{XMCD-PE} values given below are corrected by dividing by $\cos{(\Theta)}$ and plotted together with the angle dependent data in Figs.~\ref{fig:Fe11nmEuO_Tlow_XMCDPE_allAngles} and~\ref{fig:all3nmEuO_allT_XMCDPE_allAngles}. To the \emph{un}corrected values, we will in the following refer as \mbox{XMCD-PE$_{\text{meas}}$}. All samples exhibit a strong Eu-related \mbox{XMCD-PE}. For the two bare EuO films, it is noteworthy that the absolute value of the thicker film is about 1.5 times larger than that of the thinner film ($59{.}7$\% vs. $41{.}5$\%), indicating a reduced magnetic interaction in the thin film. This might be attributed to size effects, i.e. under-coordination.\\
	When interfaced with $4\,$nm Fe, the $11\,$nm EuO film exhibits an Eu-related \mbox{XMCD-PE} of the same sign as obtained for bare EuO. For later discussion, we note that the absolute value ($52{.}8$\%) is slightly reduced compared to bare EuO. Not surprisingly, also a strong Fe-related \mbox{XMCD-PE} is present.\\
	The results change completely for the \mbox{XMCD-PE} obtained from the \emph{thin} EuO film with a $4\,$nm Fe overlayer: Also here, a strong \mbox{XMCD-PE} related to both Eu and Fe is present, but their signs are inverted compared to the sample with the thick EuO film. From this observation we can draw the following conclusions: i) The magnetizations of the two magnetic layers are anti-parallelly aligned, i.e. there is a magnetic coupling between the EuO and Fe layers and it is of AFM nature. ii) In the sample with the thin EuO film, the Fe layer is dominant and dictates the magnetization direction of EuO, while roles are reversed in the sample with the thick EuO film.\\
	Taking into account the absolute values of the \mbox{XMCD-PE}, we find that the Eu-related signal ($18{.}9$\%) is reduced by a factor of about 2 compared to the bare EuO reference. This is a much stronger reduction than for the thick EuO films. Probably, the reduced Eu-related \mbox{XMCD-PE} is the consequence of two competing effects: the intrinsic tendency of EuO to align with the applied field against the AFM coupling with the dominant Fe overlayer. Possibly, this competition also causes the slight reduction of the EuO magnetization with respect to the reference sample in the case of thick EuO for which the interface effect is smaller compared to the bulk signal. The absolute value of the Fe-related \mbox{XMCD-PE} is larger for the sample with thin EuO ($41{.}3$\%) compared to thick EuO ($35{.}3$\%). This could also be explained within the picture of competing effects, where for thin EuO, Fe is magnetized in the direction of the external field, while for thick EuO the remanent state of Fe is caused by the AFM proximity coupling, which is, however, too weak to push the Fe layer up to its intrinsic remanent magnetization.\\
	For thin EuO interfaced with $4\,$nm Co instead of Fe, we observe clear Eu and Co related \mbox{XMCD-PE} signals. Again, the Eu-related \mbox{XMCD-PE} is inverted with respect to the bare EuO reference, indicating AFM coupling of the layers also for this sample. Interestingly, the absolute value related to Eu ($40{.}6$\%) is matching the bare EuO reference. From this, we conclude that Co compared to Fe is the more dominant partner for EuO. A possible explanation will be discussed below.\\
	Increasing the temperature to $T_{\text{high}}\,\approx\,80\,$K, the Eu-related signal for thick EuO interfaced with Fe is weak and hardly distinguishable from noise (as discussed later), while for the thin EuO films interfaced with both Fe or Co we observe a reduced but clear Eu-related XMCD-PE (Fe/EuO: $3{.}1$\%, Co/EuO: $3{.}9$\%). Hence, the values for both samples are similar at $T_{\text{high}}$ which is surprising because at $T_{\text{low}}$ they differ by a factor of larger than two. The smaller relative decrease with temperature for Fe/EuO allows for the assumption that the proximity effect more effectively tunes the EuO magnetism at the Fe/EuO interface than at the Co/EuO interface. This could also explain the observation of Fe being the less dominant partner for EuO than Co (see above): If the EuO magnetism is more efficiently tuned by Fe, EuO more strongly competes Fe while being magnetized. Hence, in the remanent state a reduced magnetization of the EuO layer remains. On the other hand, due to the efficient tuning, the magnetization of EuO remains more robust against the increase of temperature to $T_{\text{high}}$.\\
	In summary, our XMCD-PE data recorded in normal emission confirm that the proximity effect at the 3d~FM/EuO interface can enhance the magnetic order in EuO. An enhancement occurring only in the thinner EuO films would suggest that interfaced EuO films even thinner than those studied in this work may have a magnetic order at $T\,\geq\,$RT, in line with an earlier study~\cite{poulopoulos_induced_2014}. Fe seems to tune the magnetism of EuO more efficiently than Co.
		\begin{figure*}[tbh]
		\centering
		\hspace*{-1mm}
		\includegraphics[width=2.00\columnwidth,clip]{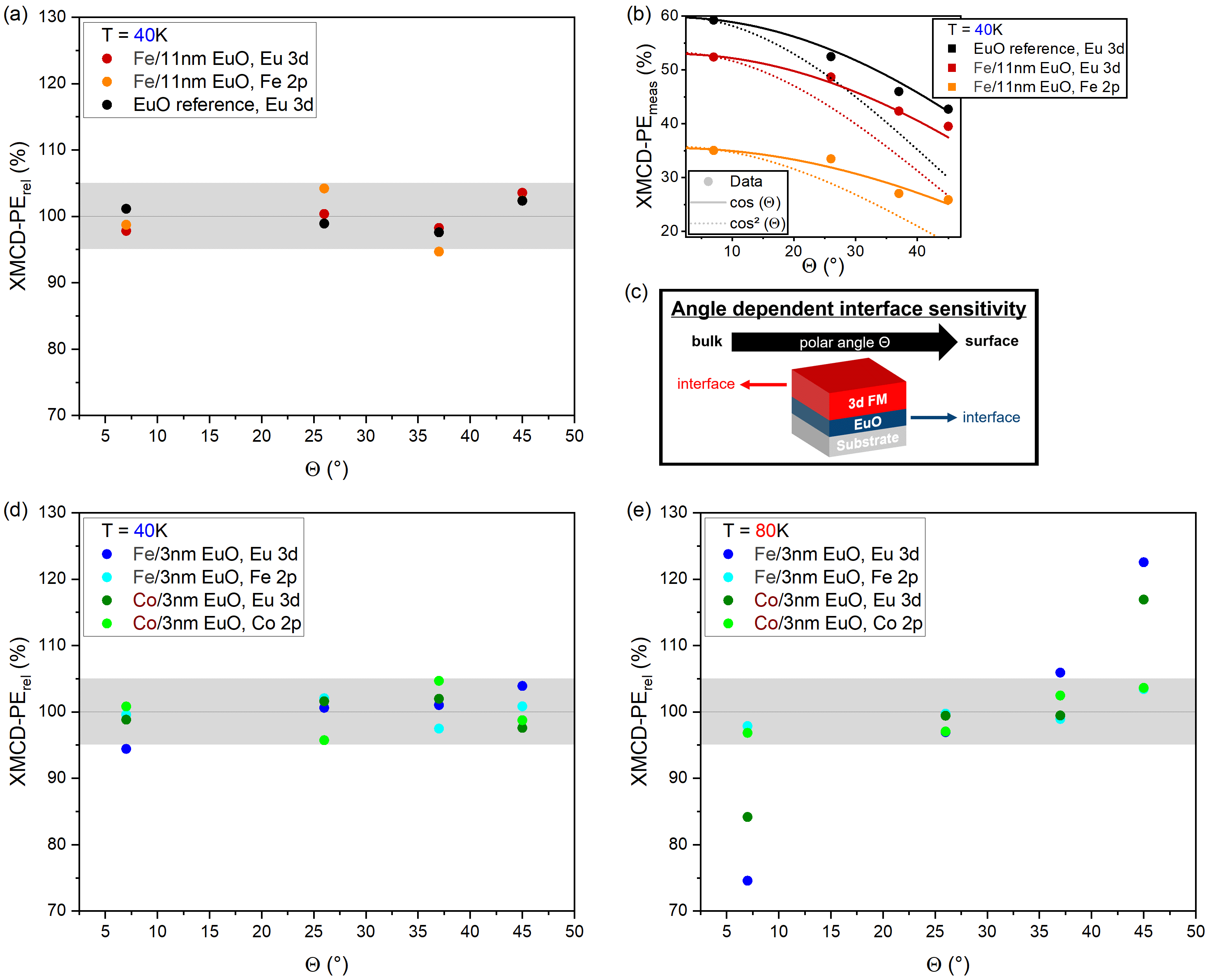}
		\caption{XMCD-PE$_{\text{rel}}$ vs. electron emission angle $\Theta$ of (a) $11\,$nm EuO w/ and w/o Fe capping, (d) $3\,$nm EuO with Fe and Co capping at $T_{\text{low}}$ and (e) at $T_{\text{high}}$. In (b), the XMCD-PE$_{\text{meas}}$ of the $11\,$nm EuO samples are plotted vs. $\Theta$. The full (dashed) reference lines which indicate a $\cos{(\Theta)}$ ($\cos^2{(\Theta)}$) dependence visualize that XMCD-PE$_{\text{meas}}$ is proportional to $\cos{(\Theta)}$. As sketched in (c), increasing \mbox{XMCD-PE$_{\text{rel}}$} with increasing $\Theta$ reveal an enhanced (reduced) magnetic moment of EuO (Fe or Co) at the interface with respect to bulk EuO.}
		\label{fig:XMCD_rel_vs_Theta}
	\end{figure*}
\subsection{\XMDP~of 3d~FM/EuO interfaces}
	The \mbox{XMCD-PE$_{\text{meas}}$} obtained under the four polar angles $\Theta$ from the bare and Fe-covered $11\,$nm EuO film at $T_{\text{low}}$ are shown in Fig.~\ref{fig:Fe11nmEuO_Tlow_XMCDPE_allAngles}. The respective plots referring to the samples with a $3\,$nm thin EuO film at both $T_{\text{low}}$ and $T_{\text{high}}$ are shown in Fig.~\ref{fig:all3nmEuO_allT_XMCDPE_allAngles}. The absolute value of the \mbox{XMCD-PE$_{\text{meas}}$} of both Eu and 3d~FM decreases with increasing $\Theta$ as expected for homogeneously magnetized films with \Mag~being oriented in-plane, as depicted in Fig.~\ref{fig:SketchSetup}. Due to the reduced photoemission intensity in off-NE geometry, the smaller $\Theta$, the larger the signal-to-noise ratio~(SNR). This effect was partially compensated by increasing the integration time for larger angles, especially for the energy range around the Eu~3d$_{5/2}$ core level as this is most relevant for the Eu-related \mbox{XMCD-PE}.
	
	For analyzing the depth dependent magnetic order, all \mbox{XMCD-PE$_{\text{meas}}$} values are divided by $\cos{(\Theta)}$ to correct for the deviation from a parallel alignment of beam direction and \Mag. The $\cos{(\Theta)}$ dependence of \mbox{XMCD-PE$_{\text{meas}}$} can be clearly seen in Fig.~\ref{fig:XMCD_rel_vs_Theta}~(b). For each sample and temperature, we calculate relative \mbox{XMCD-PE}s as,
		\begin{align}
        \label{Eq_RelXMCD-PE}
        \text{XMCD-PE}_{\text{rel}}(\Theta)= 100\% \times \frac{4\,\times\ \text{XMCD-PE}(\Theta)}{\sum_{\Theta'}\,\text{XMCD-PE}(\Theta')} \, ,
    \end{align}
	and plot them as a function of $\Theta$, Figs.~\ref{fig:XMCD_rel_vs_Theta}~(a), (d), (e). For the samples with $11\,$nm EuO at $T_{\text{low}}$, there is no $\Theta$ dependence and all values scatter within $\pm 6\%$ around the average, Fig.~\ref{fig:XMCD_rel_vs_Theta}~(a). This also holds for the $3\,$nm EuO films with Fe and Co overlayer, Fig.~\ref{fig:XMCD_rel_vs_Theta}~(d). However, a trend indicating an increasing Eu-related \mbox{XMCD-PE}$_{\text{rel}}$ with increasing $\Theta$ could be identified for the Fe-covered EuO film. This would reflect a larger degree of magnetic order of EuO in the vicinity of the Fe/EuO interface. Note that a non-zero out-of-plane component of the EuO magnetization could also contribute to the observed angle dependence.
	\begin{figure}[tbh]
		\centering
		\hspace*{-1mm}
		\includegraphics[width=0.9\columnwidth,clip]{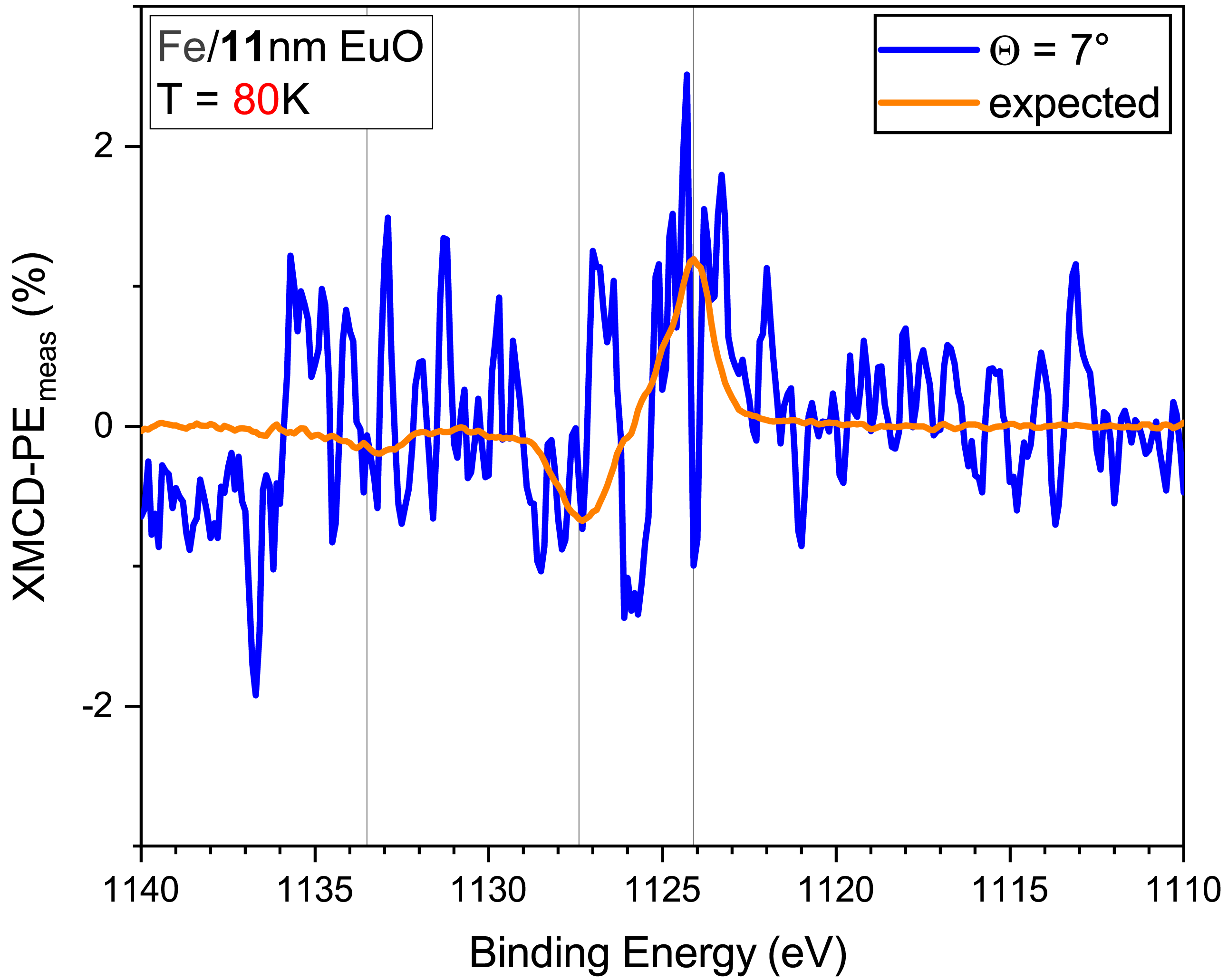}
		\caption{XMCD-PE$_{\text{meas}}$ of the Eu 3d core level obtained at $T_{\text{high}}$ and $\Theta\,=\,7^\circ$ from the Fe capped $11\,$nm EuO film (blue) and a reference signal depicting the expected angle dependence of XMCD-PE$_{\text{meas}}$ (orange). Despite the measured data is noisy, comparison with the reference signal (minimum, maximum) suggests that there might be a non-vanishing signal.} 
		\label{fig:Fe11nmEuO_Thigh_XMCDPE_7deg}
	\end{figure}	
	Clear evidence for a strong depth dependence of the magnetic proximity effect on EuO arises from the data of the thin interfaced EuO films at $T_{\text{high}}$, see Fig.~\ref{fig:XMCD_rel_vs_Theta}~(e). The Eu-related \mbox{XMCD-PE}$_{\text{rel}}$ of both samples drastically increases with increasing $\Theta$. This finding suggests: The proximity-enhanced magnetic order is a pure interface effect of limited range -- and as such a new finding for magnetic proximity effects at 3d~FM/EuO interfaces.\\
	The $\Theta$ dependence is stronger for the Fe capped film as compared to the Co capped. This supports the assumption that there could be a $\Theta$ dependence of the Eu-related \mbox{XMCD-PE}$_{\text{rel}}$ at $T_{\text{low}}$ at the Fe/EuO interface.\\
	At $T_{\text{high}}$, there also appears to be a small but non-zero increase of both the Fe and Co related \mbox{XMCD-PE}$_{\text{rel}}$ with increasing $\Theta$. This would be the result of a reduced in-plane magnetization of the 3d~FM overlayers in the vicinity of the 3d~FM/EuO interface or of a small but non-zero out-of-plane component of \Mag.
	For the $11\,$nm EuO film interfaced with Fe at $T_{\text{high}}$ and $\Theta\,=\,7^\circ$, we observe the signal shown in blue in  Fig.~\ref{fig:Fe11nmEuO_Thigh_XMCDPE_7deg}. From a bare data, a signal beyond noise can hardly be identified. We plotted the measured signal together with a scaled one obtained from a measurement at $T_\text{low}$ (orange in Fig.~\ref{fig:Fe11nmEuO_Thigh_XMCDPE_7deg}). The scaled signal reflects the expected energy dependence of XMCD-PE$_\text{meas}$. Comparison of the signal obtained at $T_\text{high}$ and the expected one suggests that also in the $11\,$nm thick EuO film with Fe overlayer there might be a non-vanishing interfacial magnetic order  above the bulk \TC~of EuO. However, a clear statement can not be made and measurements with higher integration time, i.e. better statistics, are required. Analysis of the XMCD-PE at $\Theta\,=\,45^\circ$ is already hampered by the low SNR of the bare spectra. Absence of persisting interfacial magnetic ordering would imply that the proximity effect is not only depth but also thickness dependent.

	
	

\subsection{Atomistic spin dynamics}
\begin{figure}[tbh]
    \centering
    \includegraphics[scale=0.4]{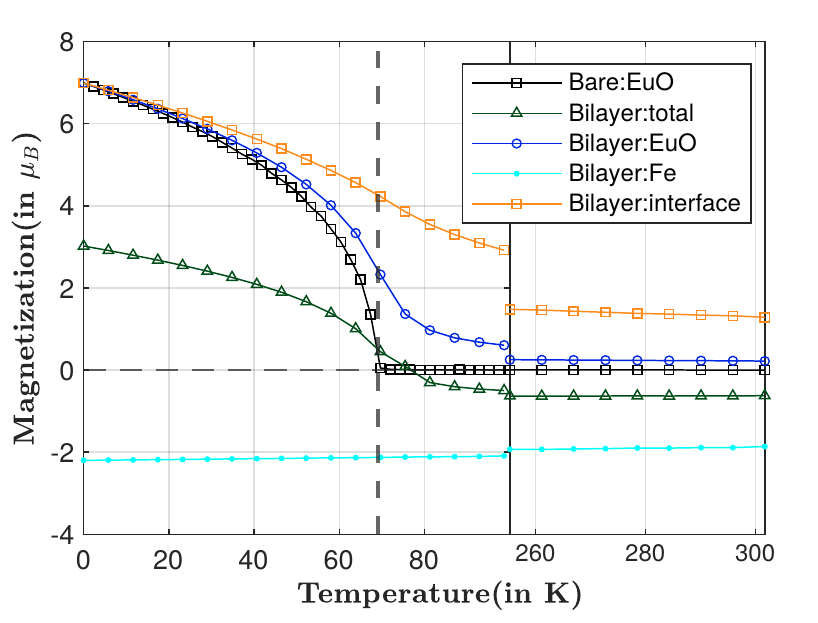}                                
    \caption{Magnetization curve of bare EuO with a thickness of 3nm, compared to the total magnetization of a EuO/Fe bilayer, where the latter magnetization is split into the EuO layer, the Fe layer, and the interface monolayer of EuO, which remains magnetized up to room temperature due to the proximity effect from Fe. From the total magnetization we find a compensation point at about $T_{\mathrm{comp}} =   75$K.}
    \label{fig:magnetization_curve}
\end{figure}
From our simulations we obtain the time evolution of the magnetization of our model, which - via a time average - turns into a thermal equilibrium magnetization. The magnetization can be split into different contributions, from the EuO, the Fe but also that from the interface monolayer of the EuO using the following temporal average:
\begin{align}
\label{7}
 m_j(T) = \mu_{B,j}\left\langle|S_j(t,T)|\right\rangle  
\end{align}
Varying the temperature we extract the magnetization curve as shown in Figure \ref{fig:magnetization_curve}. 
Here we compare simulations of a 3nm pure EuO thin film with a bilayer of EuO(3nm)/Fe(4nm). The magnetization of the latter is further split into its above mentioned parts.
Interestingly, only the pure EuO film shows a clear phase transition with a Curie temperature of about 69 K, as expected. In the bilayer, the magnetization of EuO remains always finite in the shown temperature regime, indicating the surpression of the phase transition. This is due to the fact that the interface monolayer of the EuO remains always magnetized because of its exchange coupling to the Fe layer with much higher Curie temperature (not shown). In a thin EuO layer, the average magnetization remains, consequently, also finite, even up to room temperature. 

Furthermore, the total magnetization curve shows that the bilayer with its antiferromagnetic coupling can also be interpreted as a synthetic ferrimagnet, EuO building one sublattice and Fe/Co layers the other. Over all, we find a compensation point (where the total magnetization switches sign) above the bulk \TC~of EuO. Below the compensation temperature, the total magnetization is dominated by EuO, above the compensation point the Fe sublattice dominates the total magnetization. 
\begin{figure}[tbh!]
    \centering
    \includegraphics[scale=0.5]{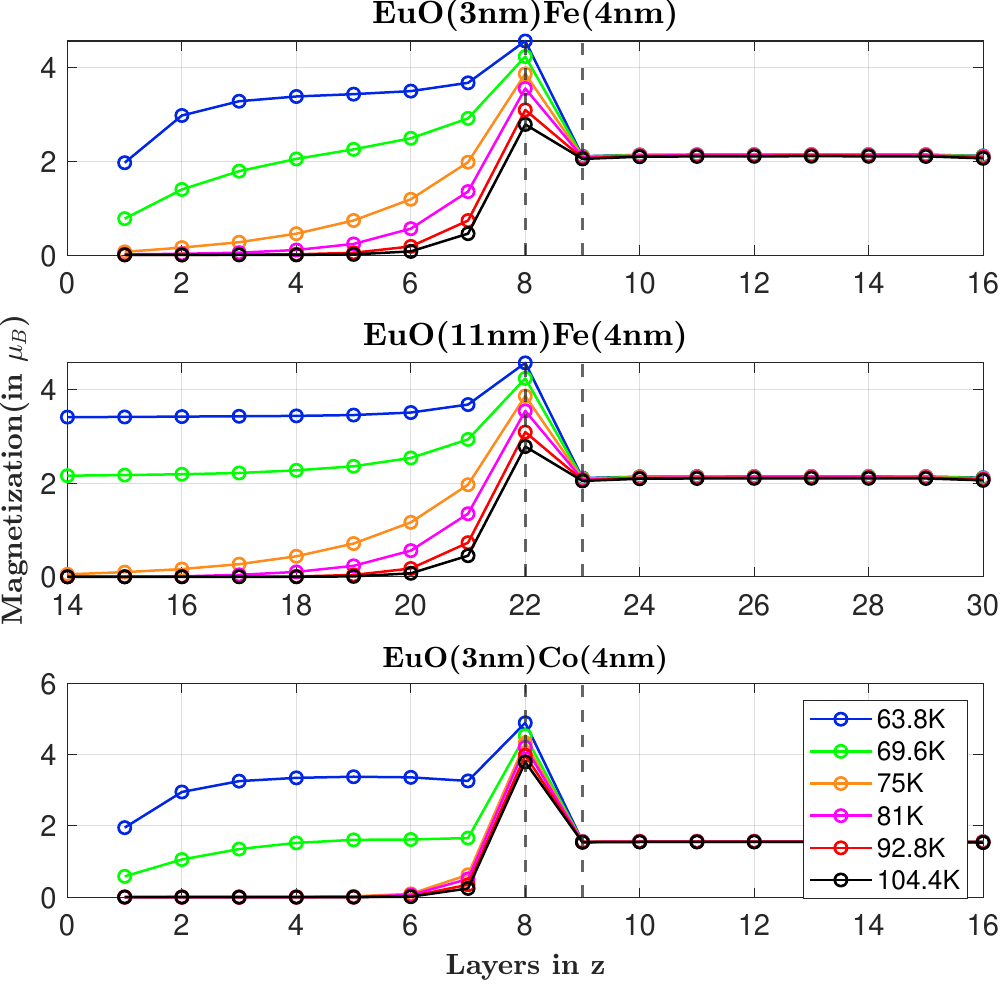}
    \caption{Spatially resolved magnetization of the samples at certain temperatures around the bulk $T_C$ of EuO. The layers to the left of the dotted lines are EuO and to the right are Fe or Co, and the dotted lines refer to the corresponding interfacial layers. It shows the dependence of the proximity effect on temperature of the sample, and the type of interface}
    \label{fig:magnetization_components_all}
\end{figure}

To investigate the nature of the proximity effect further, in Figure \ref{fig:magnetization_components_all} we present magnetization profiles for three different bilayers and for different temperatures around the Curie temperature of bare EuO. 
The most interesting feature captured in this figure is the spatial range of the proximity effect, which can be quantified as a length-scale, varying with the temperature but is - as far as possible - independent of the EuO thickness. This length-scale is maximal for 75K, slightly above the Curie temperature of EuO, indicating that it is connected to a critical phenomenon. 

The first monolayers of the EuO layer  show a reduced magnetization due to surface effects. In our simulations, we use open boundary conditions leading to fewer interactions between neighboring spins, and, hence, less magnetic order, which is clearly visible at 63.8 and  69.6K. 

Finally, we should mention that for our simulations, the quality of the interface might play an important role.  In our spin model, EuO and the 3d-transition metals are only separated via their lattice structure and the strength of their exchange interactions. EuO and Co are ordered FCC leading to a smooth interface and has more but weaker interactions whereas Fe is BCC, leading to a rough interface and has lesser but stronger interactions. Here the interfacial coordination number per atom from the EuO is not important but the coupling strength is, and that is purely an estimation(for our simulations we take this as the average of $J_{EuO}$ and $J_{Fe/Co}$) since the exact values for the interfacial couplings are not known for such systems. In case of EuO/Co bilayer, shown in the last subplot of this figure, only 2 monolayers are magnetized, which is approximately 6 monolayers in case of EuO(both thin and thick)/Fe bilayers.
It can also be predicted that if the sample thickness is at a length scale comparable to that of the proximity effect then we have all layers magnetically ordered, leading to the entire film still ordered above the bulk $T_C$. 

\section{Discussion}
      
  In our study, we analyze the depth profile of the magnetic proximity effect at a 3d~FM/Eu monochalcogenide interface. 
We find a depth dependence both on the 3d~FM side and on the EuO side of the interface; and the results also show how these depth profiles depend on the temperature. Our observation of an enhanced magnetic moment at the EuO interface challenges the result of a recent study claiming EuO is robust to proximity  effects~\cite{averyanov_probing_2019}. We also find that our bilayer behaves as a synthetic ferrimagnet with a compensation temperature above the $T_C$ of EuO due to the uncompensated interfacial coupling.
 
 \label{sec:R3:discussion}
	\begin{figure}[tbh]
        \centering
    	\hspace*{-1mm}
        \includegraphics[width=0.9\columnwidth,clip]{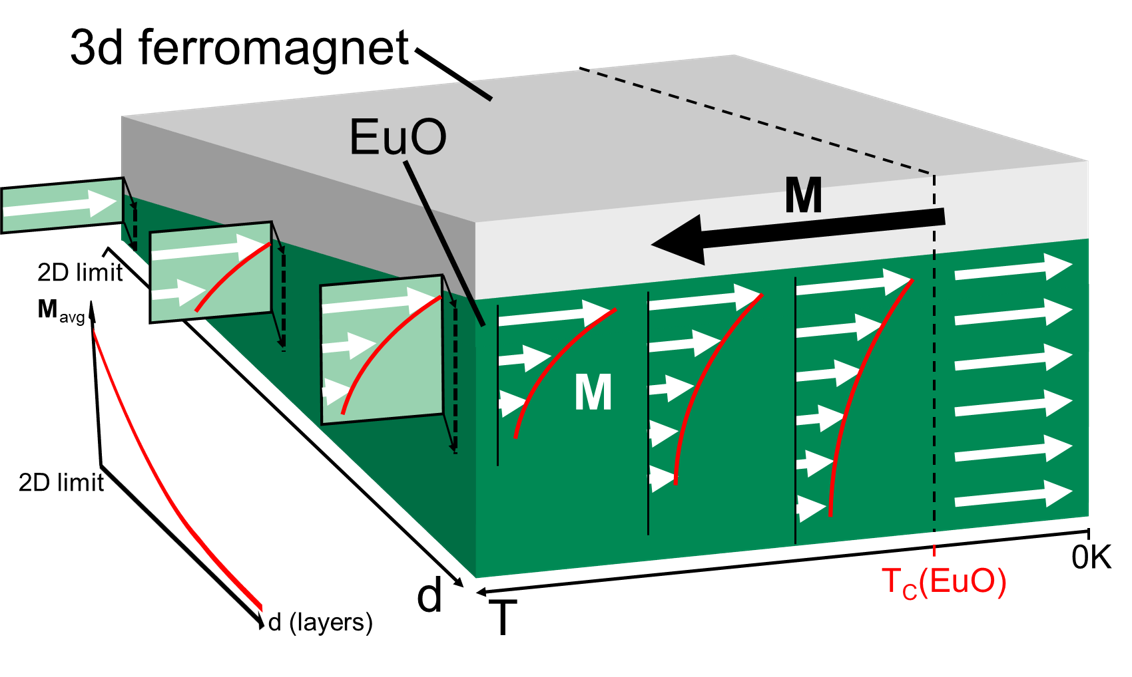}\quad
		\caption{Sketch of the key findings: (i) We confirmed an AFM coupling between a 3d~FM overlayer and EuO film. (ii) Above the bulk Curie temperature of EuO, magnetic order persists due to a proximity effect. (iii) The spatial range of this proximity effect is finite and is highest at a temperature greater than $T_C$. (iv) The proximity effect is independent of the film thickness. (v) Due to the finite range of the proximity effect, approaching the 2D limit leads to a higher total magnetization of EuO. Overall, these facts turn the EuO/Fe bilayer into a 2D ferrimagnet.} 
		\label{fig:GraphicalAbstract}
	\end{figure}
	
As explained above, two coupled layers that share one order parameter cannot have two separate phase transitions. Consequently, there is no separate phase transition in the EuO in addition to the one of the transition metal. However, looking at the layer resolved magnetization in the samples as shown in Figure \ref{fig:magnetization_components_all}, we understand that significantly enhancing the magnetic order of EuO over the full thickness by a proximity effect is only possible in the ultrathin film limit because the enhanced magnetic order is a localized effect, mostly at the interface of EuO with the 3d~ferromagnet.
  
Experimentally, we revealed the strong depth dependence of the proximity effect for interfaced thin EuO films, especially at $T_{\text{high}}$, proving that the proximity induced magnetic order is short-ranged and comprises a few monolayers only, which also varies with temperature.
Such a behavior was reported earlier in studies on EuS heterostructures, where the results were discussed in the framework of a model calculated for FeF$_\text{2}$/CoF$_\text{2}$ multilayers~\cite{carrico_phase_1992}. It shows that below the Néel temperature of FeF$_2$ and independent of the CoF$_\text{2}$ thickness, its interface layers are polarized due to the proximity to FeF$_\text{2}$. From our measurement on a thick EuO film with Fe overlayer at $T_\text{high}$, we cannot conclusively state whether there is interfacial magnetic order in EuO persisting above its bulk \TC. The absence of magnetic order would imply that the enhancement of magnetic order depends on the EuO film thickness. 
The simulations strongly support 
that magnetization persists at the interface at $80\,$K and the length scale of this proximity effect resembles the thinner EuO with Fe overlayer, since it is related to the interfacial exchange coupling. Hence we conclude that the proximity effect directly at the interface  is independent of thickness. 

From our results, we can draw conclusions about another phenomenon that may seem counterintuitive at first glance. To describe this feature, we first define a temperature, \TSM, below which the average magnetization of the EuO film remains considerable, i.e. it exceeds a certain value. From our findings we can conclude that the thinner the EuO film, the stronger the proximity-induced enhancement of \TSM~--~which contradicts the intuition that the magnetization of a thinner film is reduced due to undercoordination. To understand this phenomenon, consider that for a finite range of magnetic order amplified by proximity, which is independent of the EuO film thickness, a thickness-dependent increase in \TSM~is possible.
A temperature-dependent range of the proximity effect would play a significant role for enhancing \TSM~the more effective the thinner the EuO film is. This yields the picture of an \emph{indirect} thickness dependence of \TSM: The range of the proximity effect decreases beyond a certain critical temperature ($T>T_{C(EuO)}$) where this length-scale is maximum. For a thicker film a smaller ratio of the layers will exhibit a significant proximity-enhanced magnetic order at higher temperature. Therefore, the average magnetization will drop faster with temperature the thicker the EuO film is, see Fig.~\ref{fig:GraphicalAbstract}.

While we have demonstrated that the enhancement of the proximity-induced magnetization is short-ranged, the antiferromagnetic correlation between the 3d~FM and EuO layers turns out to be not. This fact makes the bilayer a synthetic ferrimagnet with a compensation point above the bulk $T_C$ of EuO, as shown in Figure \ref{fig:magnetization_curve}. For temperatures $T < T_{\mathrm{comp}}$, EuO has higher magnetic moment than Fe or Co and vice versa above the compensation temperature.

The magnetization of the 3d~FM overlayer seems to be slightly reduced in the vicinity of the 3d~FM/EuO interface, see Fig.~\ref{fig:XMCD_rel_vs_Theta}~(e). However, since the angle dependence is only observed at $T_{\text{high}}$, we exclude chemical changes, e.g. oxidation of the 3d~FM at the interface, as the cause. Instead, this effect is probably due to the fact that the 3d~FM interface  monolayers experiences two types of interactions, strong positive with the bulk of Fe whereas weaker negative with EuO, thus decreasing the magnetization in the FM monolayer at the interface, as illustrated in Figure \ref{fig:GraphicalAbstract}. \\
In our simulations, we find finite  magnetization in the interfacial EuO layers for temperatures up to room temperature seen in Figure \ref{fig:magnetization_curve}. This suggests a two-dimensional magnetized EuO interface state that can be controlled via exchange coupling, temperature and sample thickness. However, in contrast to the studies on EuS-based heterostructures focussed on magnetic proximity effect~\cite{fumagalli_exchange-induced_1998,lewitz_proximity_2012,pappas_direct_2013,poulopoulos_induced_2014,goschew_verification_2016}, which show ferromagnetism at RT, we studied a bilayer instead of a multilayer structure. We expect the induced magnetization and, hence, the enhancement of the magnetic order in EuO to be much stronger in a multilayer structure, as the proximity effect acts on the EuO layers from both sides. Based on our findings on the 3d~FM/EuO proximity effect, we therefore assume that further enhancing magnetic order in EuO is possible in two ways: by reducing the EuO thickness and/or sandwiching EuO between two 3d~FM layers. Moreover we conclude that in such magnetic heterostructures, spin-spin interaction is responsible for the proximity effect, since the interfacial exchange induces magnetization in EuO.

To summarize, we studied the magnetic proximity effect at Fe/EuO and Co/EuO interfaces by applying \XMDP~ to bilayers with different EuO thickness as well as reference samples and through atomistic spin dynamics simulations. Our methods  allow for a temperature dependent analysis of the depth profile of the magnetization  inside the EuO layer. 
In the thin film limit, the EuO does not exhibit any separate phase transition, since its interface layer remains always magnetized due to the vicinity of the 3d-TM. This fact turns our bilayers into synthetic ferrimagnets with a compensation temperature above the bulk \TC \; of EuO, that may combine the easy control of net magnetization by an external field with an antiferromagnetic-like dynamics faster than ferromagnetic dynamics and the potential for high-density devices. Our findings underline that reducing the EuO thickness towards the 2D limit and interfacing the layer from both sides is a promising approach for pushing the induced magnetization of the EuO into the technologically relevant temperature range. 

	\begin{acknowledgments}
    This work was supported by the Deutsche Forschungsgemeinschaft through the International Collaborative Research Center TRR160 (Project C9) and Sonderforschungsbereich SFB 1432 (Projects B02 and B3). We acknowledge DESY (Hamburg, Germany), a member of the Helmholtz Association HGF, for the provision of experimental facilities. Parts of this research were carried out at PETRA III using beamline P22. Funding for the (HAXPES) instrument by the Federal Ministry of Education and Research (BMBF) under framework program ErUM is gratefully acknowledged.
	\end{acknowledgments}

	\bibliography{MAIN.bib}

\end{document}